\documentclass[preprint,preprintnumbers,amsmath,amssymb]{revtex4}

\usepackage{graphicx}

\begin{document}
{\sf{\large
\title{Convection onset induced by a density stratification whose unstable part is infinitely thin}
\author{N. H. Aljahdaly}
\email{nhaljahdaly@crimson.ua.edu}
\author{L. Hadji}
\email{lhadji@ua.edu}
\affiliation{The University of Alabama,
Tuscaloosa, Alabama 35487}
\maketitle}

\begin{center}\bf{Abstract}\end{center}
We consider a vertical cavity composed of two chambers separated by a retractable thermally insulated thin membrane. The upper  and lower chambers are filled with an incompressible Boussinesq fluid and maintained at temperatures $T_2$ and $T_1>T_2$, respectively by two separate heaters. Upon retraction of the membrane, the two fluid masses form an unstably stratified configuration with cold and heavy fluid overlying a warmer and lighter fluid and separated by a non-free interface across which there is a jump in the density.  
The aim of this paper is to determine the threshold conditions for convection onset and associated fluid flow patterns induced by this discontinuous density stratification. We find that the discontinuity of the density profile leads to the appearance of temperature perturbation iso-contours that have a lens shape instead of the classical oval shape and that the mixing is confined to near the location of the density jump with stagnant and isothermal fluid away from the discontinuity. We derive the flux conditions at the discontinuity interface and show that it acts like a heat sink for both the lower and upper fluid layers. Moreover, we put forth the dependence of the stability threshold parameters for convection onset on the locus of the density discontinuity. The  experimental set-up suggested in the paper can be used to test the predictions of the present theory.

\pacs{47.20.Ky, 47.55.P-, 47.54.-r}

\maketitle

\newpage

\section{Introduction}
Convection plays a central role in a myriad of natural and industrial processes such as magma flows, astrophysical and geophysical flows, materials processing, heat and mass transfer in crystalline rocks and countless man-made fluidic devices, just to name a few. Our understanding of this phenomenon stems from  studies of Rayleigh-B\'{e}nard (RB) convection \cite{1} that have been taking place since the beginning of the twentieth century. RB convection pertains to the gravitational instability that ensues when a thin fluid layer that is confined between two horizontal plates is heated from below leading to the formation of an unstable linear continuous base profile for the density. The warmer fluid is lighter than the overlying colder fluid leading to the onset of instability when the temperature difference between the two plates exceeds some threshold value. The mathematical simplicity of the base profile combined with the ease with which it can be implemented in experiments have made the RB set-up a valuable resource, the use of which has stimulated the development of related theory and experiment to further our understanding of buoyancy-driven flows. There are many other realistic settings of convection that are not well described by a reference state having a linear density profile. For example, there are physical situations where the reference state has an unstable stratification that occupies only a fraction of the total depth of the fluid as in the case of penetrative convection and convection flows induced by internal heat sources. For instance, Batchelor and Nitsche \cite{2} considered the stability of an unbounded two-dimensional fluid region that is subjected to a variety of base profiles having density stratification with zero gradient except in a centrally located region of small thickness. Although the assumed profiles are idealized, they are able to capture the main stability features that are observed in experiments involving sedimentation waves in liquid suspensions of small particles. Matthews \cite{3,4} posited a basic temperature that is a cubic polynomial in the vertical coordinate to model penetrative convection in lakes induced by solar heating. 
Simitev and Busse (SB) \cite{5} posited the following base profile for the temperature to elucidate certain features of astrophysical
turbulent convection, 
\begin{equation}
 T_B (z)=\beta(z-0.5)- \tanh(\xi(z-0.5))/\xi, \qquad z \in [0,1]
\end{equation}
For large $\xi$ values, convection takes place in a centrally located thin layer that is surrounded by stably stratified layers from above and below, thus reducing the influence of the horizontal boundaries on the stability characteristics of the convection state. Hadji {\it et al.} \cite{6} considered the case  $\xi \rightarrow \infty$ and $\beta=0$ to obtain a step function base profile to describe linear convection induced by an infinitesimally thin and unstably stratified central layer and derived stability threshold for a variety of velocity and temperature boundary conditions. The stability of this base state is being investigated in this paper when the fluid region is either unbounded in the vertical direction or has a very large aspect ratio defined as the height over width. For this purpose, we consider the vertical analog of the RB set-up, namely a fluid layer that is confined between two walls having a large or infinite vertical extent. Furthermore, we consider a base density profile having a jump discontinuity  at some prescribed location $Z_0$ and inducing a vertical unstable stratification. The objective is to uncover the influence of an infinitesimally thin and unstably stratified layer on the onset of convection in an infinitely high channel so that the development of convection is confined to take place in the vertical direction. We discovered that the stability threshold conditions for the infinitely high channel are reached whenever  the aspect ratio is $\sim 4$ are very close to those of the open cavity. Thus, the results for the closed  cavity are depicted for aspect ratios between $2$ and $4$. While the problem is formulated using temperature, the analysis applies to any diffusing element with the conservation of energy replaced by a conservation of species equation.  While in \cite{5} the main purpose of the work was not the analysis of a realistic physical model but rather to posit that some basic profile can be maintained by a combination of sources and sinks in order to be able to analyze and further our understanding of a convection phenomena for which, in the absence of such simplification, the analytical solutions are out reach. In this paper, we are modeling a physical situation, the experimental set-up of which is feasible as depicted in Fig. (~\ref{model1}).

The outline for the remainder of the paper is as follows. The mathematical formulation of the problem is presented in Section 2. In Section 3, we present the stability results for the two-dimensional cavity. The paper closes with a discussion of the results in Section 4. The large number of expressions generated during the calculations are displayed within an appendix section.

 \section{Mathematical formulation}
 
 \begin{figure}[h]
     \centering
    \includegraphics[scale=0.32]{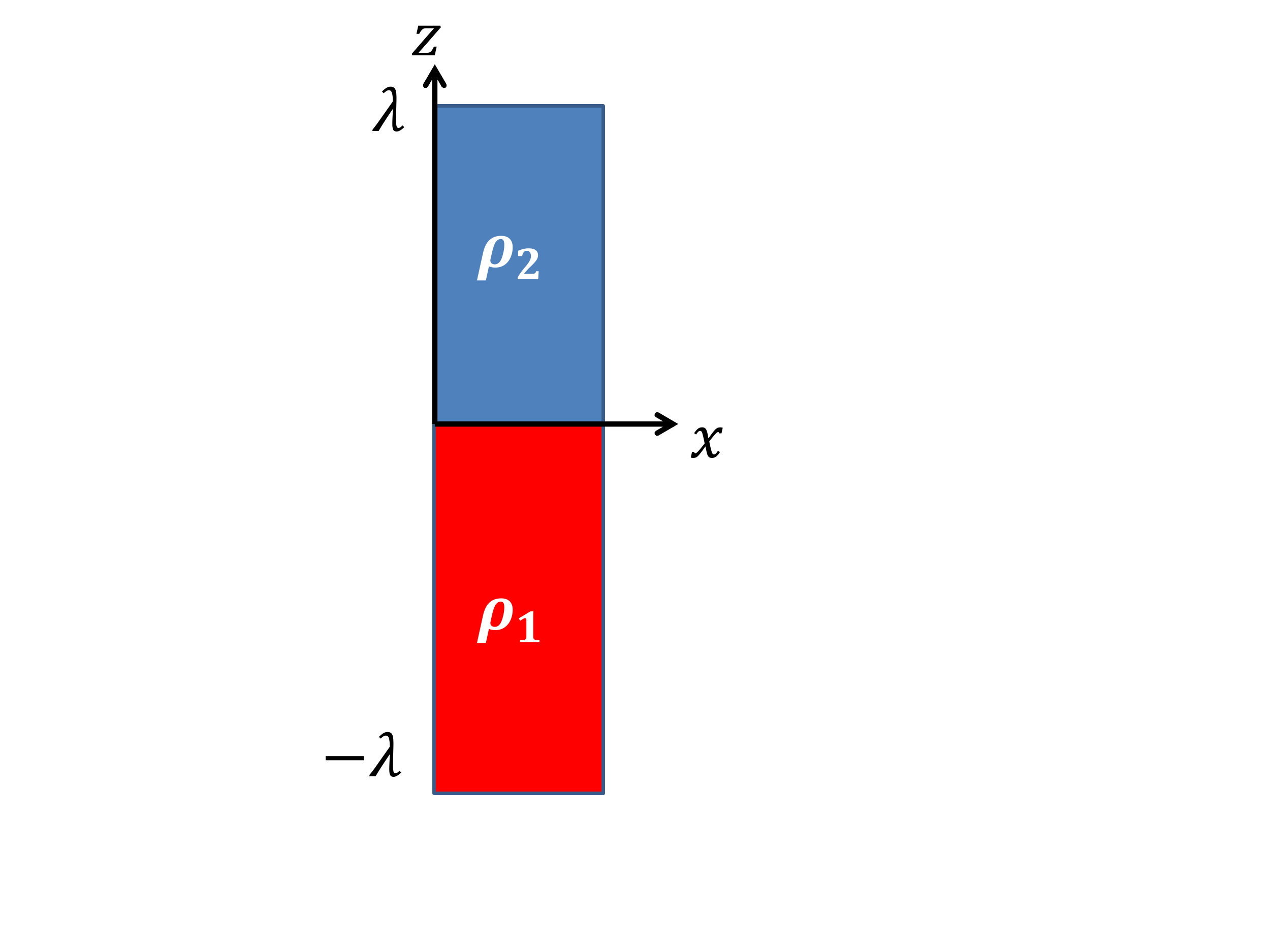} 
    \includegraphics[scale=0.32]{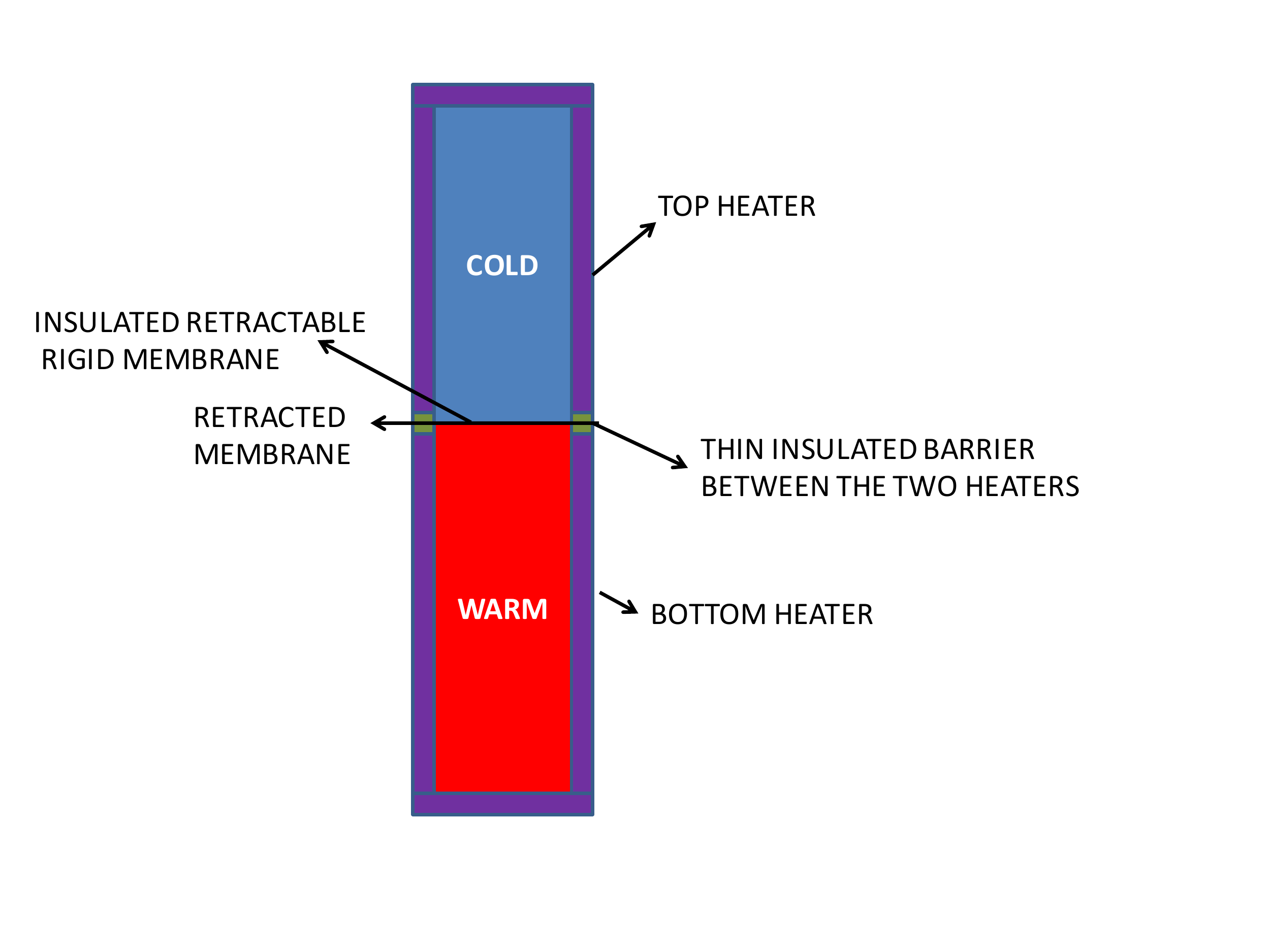}
     \caption{A schematic diagram of a fluid in a tall cavity subjected to a step function conduction profile ($\rho_1<\rho_2$) with $Z_0=0$. Unlike the Rayleigh-Taylor instability, this unstable configuration, the realization of which is described in the figure on the right, can remain stable up to a threshold density difference due to the diffusion of buoyancy. An experimental procedure is suggested wherein the upper and lower parts of the fluid mass are maintained at temperatures $T_2$ and $T_1>T_2$, respectively by top and bottom heaters. The two heaters are separated by a thin thermally insulated barrier and the two fluid parts are separated by a retractable thermally insulated membrane at $z=Z_0$. The mixing of the cold and warm fluid is achieved by the pulling of the membrane. The narrow vertical channel allows us to use a shorter membrane to minimize the mechanical disrurbance to the fluid upon its pulling.}
     \label{model1}
 \end{figure}
We consider the situation wherein the warm fluid occupies the lower region, $z < Z_0$, and the cold fluid occupies the top region, 
$z > Z_0$, as shown in Fig.\ref{model1}, leading to lighter fluid underlying a heavier one. This configuration resembles that of the Rayleigh-Taylor problem where a heavy fluid overlies a lighter one and separated by a deformable free-interface. The configuration being considered here, however, consists of the same fluid maintained at two different temperatures and occupying two regions that are separated by a non-free planar interface of zero thickness, $z = Z_0$, across which there is also the stabilizing effect of buoyancy diffusion. The base profile is described by ${\hat T_B}=T_1+(T_2-T_1) \mathcal{H}(z-Z_0)$ where $\mathcal{H}(z)$ is Heaviside function. The governing system of equations is given by \cite{7,8}
%\begin{subequations}
\begin{equation}
\frac{\partial {\mathbf{v}}}{\partial t} =-\nabla p+R\,\theta {\mathbf{k}}+ \nabla^2{\mathbf{v}} \label{eq1}
\end{equation}
\begin{equation}
Pr\,\frac{\partial \theta}{\partial t}=\nabla^2\theta+{\mathbf{v}} \cdot (DT_B(z)) \mathbf{k}\label{e2}
\end{equation}
\begin{equation}
\nabla \cdot {\mathbf{v}}=0\label{e3}
\end{equation}
%\end{subequations}
where ${\mathbf{v}}$ is the velocity vector, $\theta$ is the temperature perturbation, namely $T=T_B(z)+\theta$, where $T_B$ is the basic dimensionless temperature, $T$ is total temperature and $p$ is the deviation of the pressure from its reference state value $P_B(z)$ given by  $P_B(z)=R\,Z_0$ for $z<z_0$ and $P_B(z) = -R\,z$, for $z>Z_0$. 
The continuity equation Eq. (~\ref{e3}) is eliminated from the system formulation after introducing the general representation for a divergence-free vector field  \cite{9},
$$
{\mathbf{v}} = {\mathbf{\nabla}} \times ({\mathbf{\nabla}} \times \Phi\,{\mathbf{k}}) = {\mathbf{\Pi}}\Phi 
$$
and operating with $({\mathbf{\Pi}} \cdot) $ onto Eq. (\ref{eq1}) to obtain the following
dimensionless  linearized system, 
 \begin{equation}
 \begin{split}
\nabla^2\Phi_t+R\theta-\nabla^4\Phi&=0\\
Pr\theta_t=&\nabla^2\theta- DT_B(z)\nabla^2_H \Phi.
\end{split}
\label{basicsystem}
\end{equation}\\
 The symbol $\Phi$ represents the poloidal representation of the divergence-free velocity field, the operator $D={\partial}/{\partial z}$, $DT_B=-\delta(z-Z_0)$ where $\delta(\bullet)$ represents the Dirac delta function, $\nabla^2=\nabla^2_H+D^2$ and  $\nabla_{H}^2={\partial^2}/{\partial x^2}+{\partial^2}/{\partial y^2}$ is the horizontal Laplacian. The main eigenvalue of the problem is the Rayleigh number $R$ given by $R=(\alpha_{\theta} (T_1-T_2) d^3 g)/(k \nu)$, where $d$ is separation between the vertical walls, $\nu$ is the viscosity, $\alpha_{\theta}$ is the thermal expansion coefficient, $g$ is the gravitational constant, $k$ is the thermal diffusion coefficient. The remaining parameter is the Prandtl number, $Pr =\nu/k$, which represents the ratio of viscous to thermal diffusion.

%--------------------------------------------------------------------------------------
\section{Results}
We consider cavities that are either closed, $ \{(x,z)|\,0 \le x \le 1,\,-\lambda \le z \le \lambda \}$ or open
$ \{(x,z)|\,0 \le x \le 1,\,-\infty < z < \infty \}$.
The linear two-dimensional and time independent governing system of equations reduces to the following single boundary value problem for $\Phi$: 
\begin{equation}
\nabla^6\Phi=R\,\delta(z-Z_0)\Phi_{xx}.
\label{eigenvalue1}
\end{equation}
with the boundary conditions for the closed cavity case given by
\begin{equation*}
\begin{split}
\Phi(x,z)=&\Phi_{xx}=0, \quad {\mbox{at}} \quad x=0,1, \quad -\lambda < z < \lambda\\
\Phi(x,z)=&\Phi_z=0, \quad {\mbox{at}} \quad z=\pm\lambda, \quad 0 < x <1\\
\theta(x,z)=&0, \quad {\mbox{at}} \quad x=0,1, \quad z=-\lambda,\,\lambda.
\end{split}
\label{BC}
\end{equation*}
The same boundary conditions apply for the open cavity case with the exception that we impose that the solutions be bounded at $z=\pm \infty$. These boundary conditions pertain to boundaries that are  perfectly thermally conducting. Note that the side walls are kept at the same temperature so that the imposed heat flux acts  solely in the vertical direction..
%=====================================

In view of the form of the boundary conditions at the vertical walls, Eq. (\ref{eigenvalue1}) has solutions of the form $\Phi(x,z)=F(z)\sin(n\pi x)$. It then follows that
\begin{equation}
(-(n\pi)^2+D^2)^3F=-(n\pi)^2 R\delta(z-Z_0)F.\label{eqmethod3}
\end{equation}
 Due to the presence of the delta function term, we solve Eq. (\ref{eqmethod3}) in two separate regions, namely $F^-(z)$ for $z<Z_0$ and $F^+(z)$ for $z>Z_0$ to obtain,
\begin{equation}
 F^-(z)=(-n \pi B_1 (z+\lambda)) \cosh(n \pi (z+\lambda))+(B_1+B_2 (z+\lambda)+B_3 (z+\lambda)^2)\sinh(n \pi(z+\lambda))
\end{equation}
\begin{equation}
 F^+(z)=(-n \pi C_1 (z-\lambda)) \cosh(n \pi (z-\lambda))+(C_1+C_2 (z-\lambda)+C_3 (z-\lambda)^2)\sinh(n \pi(z-\lambda))
\end{equation}
where the $B_i$'s and $C_i$'s, $i=1,2,3$ are constants to be determined by imposing five continuity conditions for $F$ and its first four derivatives, $F^{(i)}(z)$, $i=0,1,2,3,4$, and  the jump condition, $D^5F^-(Z_0)-D^5F^+(Z_0)=-R(n\pi)^2F(Z_0)$. The latter consists of the jump condition across the discontinuity and is obtained by integrating the differential equation, Eq. (~\ref{eqmethod3}), over the interval $[Z_0-\ell,Z_0+\ell]$ and then taking the limit of the resulting integration as $\ell$ approaches $0$. These continuity and jump conditions result in a homogeneous system of linear equations for the six constants.  ${\hat D}{\bf U}={\bf 0}$, where the entries of the matrix ${\hat D}$ are depicted in the appendix section. The eigenvalue $R$ is then obtained by invoking the fact that a non-trivial solution exists if and only if det$(\hat{D})=0$. Table I. shows the results depicting the dependence of the critical Rayleigh number, $R_c$, which by the way occurs for $n=1$, on both the cavity height and the location of the discontinuity $Z_0$. The decrease of $R_c$ with $\lambda$ is attributed to the increase of the height of the region wherein convection takes place;  $R_c(Z_0)$ is found to be  symmetric function of $Z_0$, the minimum of which is attained at $Z_0=0$ with $R_c(Z_0) \rightarrow \infty$ as $Z_0 \rightarrow \pm \lambda$. The latter is an expected finding since the limits $Z_0 \rightarrow \pm\infty$ correspond to regions that are stably stratified over the whole extent of the fluid region.

\begin{table}[]
\centering
\begin{tabular}{c|c||c|c||c|c||c|c}
$\lambda=1$ & &$\lambda=2$ & & $\lambda=3$& & $\lambda=4$\\ \hline
 $Z_0$& $R_c$ &$Z_0$& $R_c$&$Z_0$& $R_c$&$Z_0$& $R_c$\\ \hline
 -0.8& 1933.1& -0.8&172.0&-0.8&165.42&-0.8&165.37\\
 -0.5& 337.81&-0.5&167.06&-0.5&165.38&-0.5&165.37\\ 
 -0.2&213.73&-0.2&165.82&-0.2&165.37&-0.2&165.37\\ 
 0&200.14&0&165.65&0&165.37&0&165.37\\
 0.2&213.73&0.2&165.82&0.2&165.37&0.2&165.37\\ 
 0.5&234.04& 0.5&166.05& 0.5&165.37&0.5&165.37\\ 
 0.8&1933.1& 0.8&172.01& 0.8&165.42&0.8&165.37\\ 
\end{tabular}
\caption{Critical Rayleigh number as function of $Z_0$ and $\lambda$.}
\label{tableL2F}
\end{table}
The temperature perturbation is then found by solving Eq. (~\ref{basicsystem}) to obtain
 \begin{eqnarray}
\theta^+(x,z)&=A^+ \sinh{(\pi(z-\lambda))} \sin{(\pi x)} \\ \nonumber
\theta^-(x,z)&=A^- \sinh{(\pi(z+\lambda))} \sin{(\pi x)}.
\label{theta0}
\end{eqnarray}
where $A^+$, $A^-$ are constants. The continuity of $\theta$ at $Z_0$ implies,
\begin{equation}
A^-=A^+\frac{\sinh{(\pi(Z_0-\lambda))}}{\sinh{(\pi(Z_0+\lambda))}}\label{theta23}
\end{equation}
Therefore, the derivatives $D\theta^+$ and $D\theta^-$ can be expressed as
\begin{eqnarray}\label{dtheta}
D\theta^+(x,z)&=&A^+\pi\cosh{(\pi(z-\lambda))}\sin{(\pi x)}\\ \nonumber
D\theta^-(x,z)&=&A^+\pi\frac{\sinh{(\pi(Z_0-\lambda))}}{\sinh{(\pi(Z_0+\lambda))}}\,\cosh{(\pi(z+\lambda))}\,\sin{(\pi x)}.
\end{eqnarray}
and upon application of the jump condition, 
\begin{equation}D\theta^+-D\theta^-=-(\pi)^2\Phi~~\text{at}~z=Z_0,\label{jumptheta}\end{equation} we solve for the constants $A^+$ and $A^-$,
%\begin{equation}
%A^+(n\pi) \cosh((n \pi) (Z_0-\lambda))-A^+\frac{\sinh(n \pi (Z_0-\lambda))}{\sinh((n \pi) (Z_0+\lambda)}(n\pi) \cosh((n \pi) (Z_0+\lambda))=-R(n\pi)^2F(Z_0).\label{theta33}
%\end{equation}
\begin{eqnarray}\label{constanttemp}
A^+&=-\pi\,F^+(Z_0)\Big(\frac{\sinh( \pi(Z_0+\lambda))}{\sinh(2\pi \lambda)}\Big)\\ \nonumber
A^-&=-\pi\, F^-(Z_0)\Big(\frac{\sinh(\pi(Z_0-\lambda))}{\sinh(2\pi \lambda)}\Big) 
\end{eqnarray}
Upon combining the equations, Eqs.(~\ref{dtheta}),  and making use of Eqs. (~\ref{constanttemp}), we obtain the following Robin type conditions for the temperature perturbation at the discontinuity corresponding to the marginal stability state at $Z_0=0$,
\begin{equation}
\begin{split}
D\theta^+=&-\pi\,\coth{(\pi\,\lambda)}\,\theta^+,\\ 
D\theta^-=&\pi\,\coth{(\pi\,\lambda)}\,\theta^-.
\end{split}
\label{flux}
\end{equation}
These relations point to a downward heat flux in the upper layer and an upward heat flux in the lower layer taking place right at the discontinuity locus. Thus, the discontinuity acts as a heat sink for both layers. Another relation  between the jump in the heat flux and the temperature perturbation is given by
\begin{equation}\label{flux2}
D\theta^+ -D\theta^-=-2 \pi\,\coth{(\pi\,\lambda)}\,\theta, \quad
{\mbox{at}} \quad z=Z_0=0
\end{equation}
which when compared to the jump condition that is obtained directly from Eq. (~\ref{jumptheta})
 \begin{equation}\label{jump}
 D\theta^+= D\theta^- - \pi^2\,\Phi, \quad {\mbox{at}} \quad z=Z_0=0
 \end{equation}
 leads to an expression relating the value of $\theta$ and $\Phi$ at the discontinuity, namely
 \begin{equation}\label{relation}
 \theta = {\pi \over 2 \coth{(\pi\,\lambda)}}\,\Phi, \quad {\mbox{at}} \quad z=Z_0=0, \quad x \in [0,1]
 \end{equation}
 This relation, Eq. (~\ref{relation}), is well verified by the graphs of $\theta$ and $\Phi$ shown in Fig. ({~\ref{Fig22}) where for 
 $\lambda =1$, we note that $\theta = 1.5649\,\Phi$ at $z=Z_0=0$.
 \begin{figure}[!ht]
	\centering
	\includegraphics[scale=0.15]{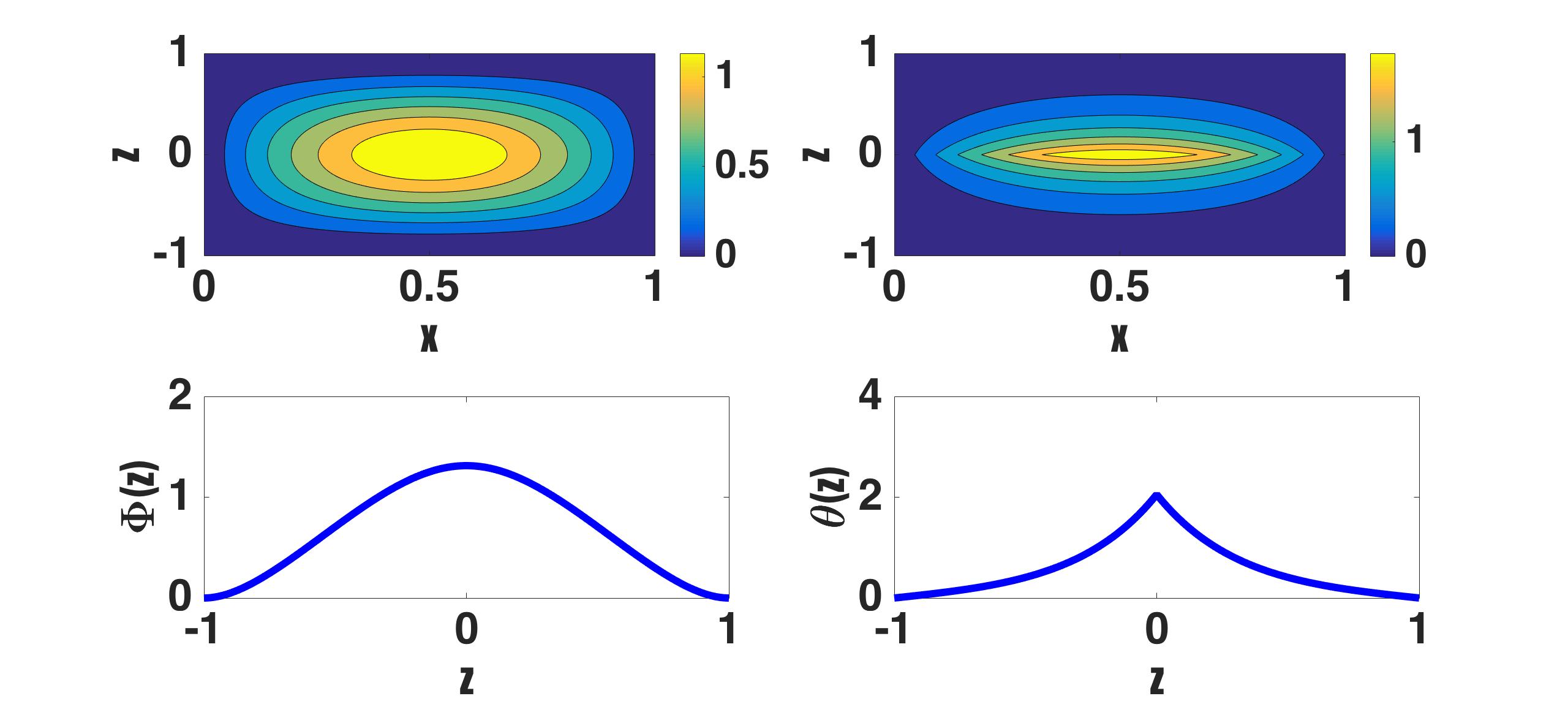}  %{twoclosed.eps}
	\caption{Plot of the streamlines (left-top), $\theta$-isotherm (right-top), velocity profile (left-bottom) and  temperature perturbation (right-bottom) versus $z$ at marginal stability for $\lambda=1$ and $x=1/2$}
	\label{Fig22}
\end{figure}

As depicted in Fig. (~\ref{Fig22}), the singularity in the temperature profile gives rise to $\theta$-iso-contours having the shape of a double pointed oval or a symmetric lens. The singularity of the solution appears at the two ends where the top and bottom arcs connect. 
\begin{figure}[!ht]
	\centering
	\includegraphics[scale=0.4]{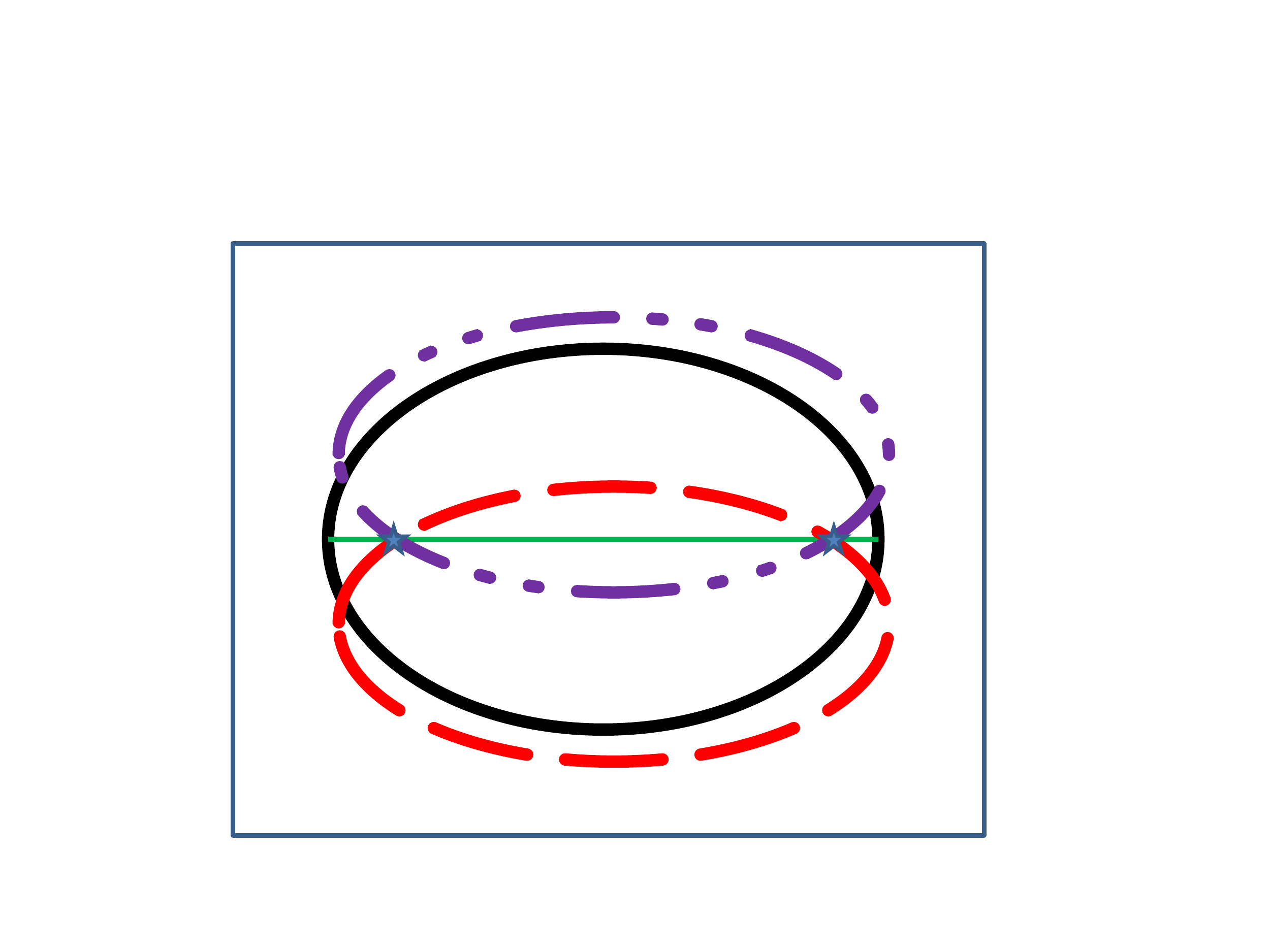}  %{twoclosed.eps}
	\caption{Sketch showing how an initial oval iso-contour for $\theta$ (continuous line) is distorted by the presence of the discontinuity at $Z_0=0$; the presence of the discontinuity acts to move the ellipse downward in the upper layer and upward in the lower region resulting in the appearance of the symmetrical lens with endpoints marked by the $\star$ symbol.}
	\label{Fig23}
\end{figure}
 Figure (~\ref{Fig23}) depicts a schematic diagram to explain the effect of the presence of the discontinuity on the $\theta$-iso-contours. In the absence of the discontinuity at $Z_0=0$, the $\theta$-iso-contours are expected to have approximately an oval shape which can be approximated by $\theta(x,z) \approx x^2/p^2 + z^2/q^2$, where the values of $p$ and $q$ are determined by the value of a specific contour level. For this symmetric shape, the heat flux $D\theta$ is continuous at $z=0$. The presence of the discontinuity at $z=0$ acts to move this symmetric ellipse away from $z=0$ to have instead
\begin{eqnarray}
\theta^+(x,z) = {x^2 \over p^2} + {(z-c^+)^2 \over q^2}, \quad z>Z_0=0,\\ \nonumber
\theta^-(x,z) = {x^2 \over p^2} + {(z+c^-)^2 \over q^2}, \quad z<Z_0=0,
\end{eqnarray}
where due symmetry considerations between the top and bottom layers, $c^-=c^+$. The jump in the heat flux at $z=0$ is then given by
$$
D\theta^+ - D\theta^- = {-2 \over q^2}(c^+ + c^-) 
$$
and on invoking Eq. (~\ref{jump}), we find that the amount by which the symmetric ellipse is moved upward and downward by the discontinuity is given by $c^+ = c^- = \pi^2 F(0)/4$. 
 
%==================

For the case of the vertically unbounded two-dimensional cavity, which is obtained from Eqs. (~\ref{eigenvalue1}) and boundary conditions corresponding to the limit $\lambda \rightarrow \infty$, we have 
\begin{eqnarray}
\Phi^-(x,z)=&F^-(x,z)\sin(n\pi x)  \quad {\mbox{for}} \quad z<Z_0\\
\Phi^+(x,z)=&F^+(x,z)\sin(n\pi x) \quad {\mbox{for}} \quad z>Z_0\\
F^-(x,z)=&(C_1 +C_2 z+C_3 z^2) e^{n\pi z}\\
F^+(x,z)=&(B_1 +B_2 z+B_3 z^2) e^{-n\pi z}
\end{eqnarray}
where $n=1,2,....$. By imposing the continuity of $F$ and its first four derivatives along with the jump condition, we obtain a homogeneous system of six linear equations, the solution of which yields the stability threshold values. We find that the marginal stability state occurs for $n=1$ and that $R_c =165.37$ independently of the value of $Z_0$, a result that is consistent with those of the closed cavity in the limit  $\lambda$ approaching infinity. The temperature perturbation is then given by
\begin{eqnarray}
\theta^+(x,z)&=A^+ e^{-\pi z}\,\sin{(\pi x)} \\ \nonumber
\theta^-(x,z)&=A^-  e^{\pi z}\,\sin{(\pi x)}
\label{theta2}
\end{eqnarray}
\begin{eqnarray}
A^+&=\frac{\pi}{2}e^{\pi Z_0}\,F^+(Z_0)\\ \nonumber
A^-&=\frac{\pi}{2} e^{-\pi Z_0} \,F^-(Z_0)
\label{constanttemp2}
\end{eqnarray}
Both the streamlines and $\theta$-isotherms bear a strong resemblance to those of the closed cavity case near $Z_0$ where the flow and  temperature variations are most intense, with nearly motionless and isothermal regions away from $Z_0$. In the limit $\lambda \rightarrow \infty$, convection takes place in an even narrower region around $Z_0$ surrounded by a stagnant and isothermal mass of fluid. Furthermore, we note that for the open cavity, the relation between the values of $\theta$ and $\Phi$ at $Z_0$ is given by
  $\theta(Z_0) = \pi\,\Phi(Z_0)/2$, a relation that is well verified by the plots in Fig. (~\ref{rcz0open}), where we note that $\theta =
  1.5708\,\Phi$ at $z=0$.

\begin{figure}[!ht]
\centering
\includegraphics[scale=0.15]{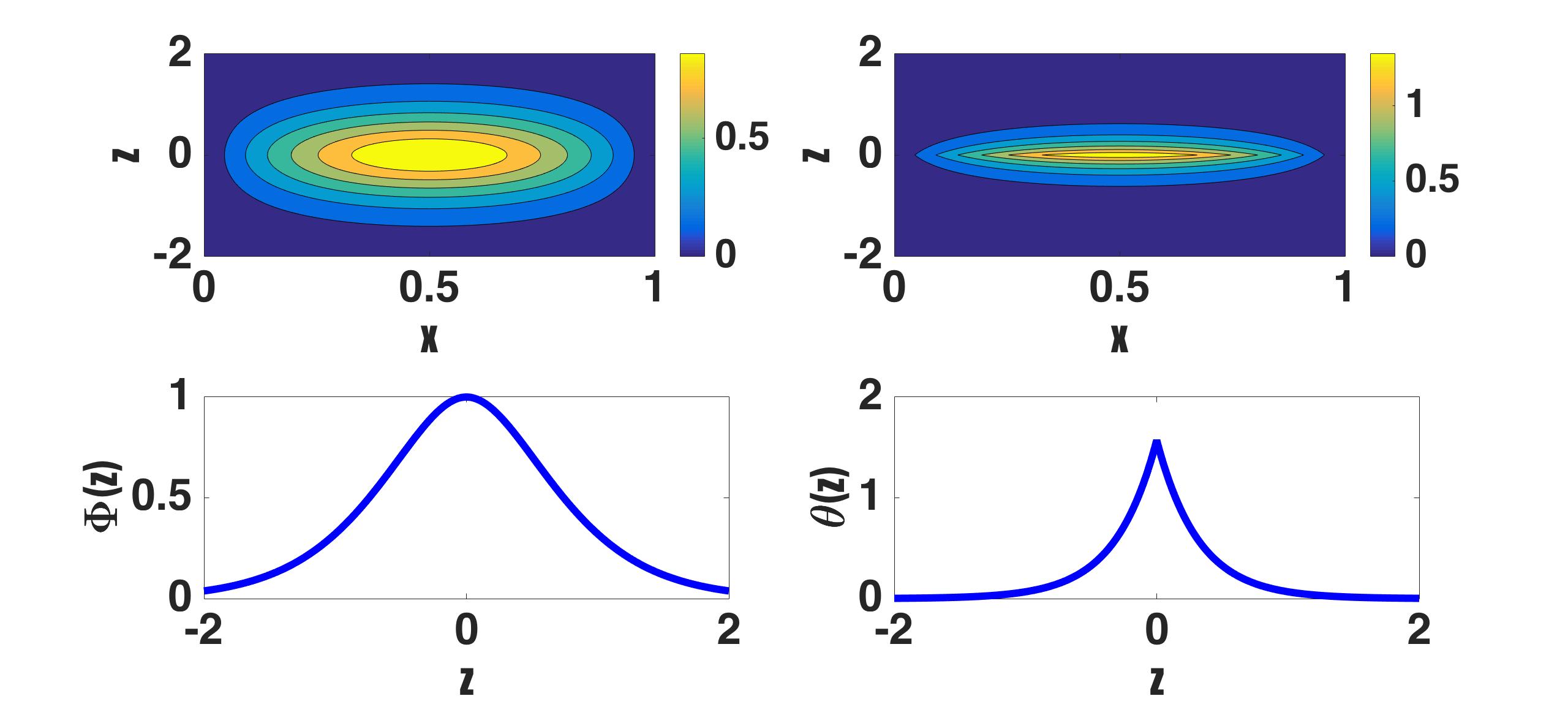}  %{twoopen.eps}
\caption{Plot of the streamlines (left-top), $\theta-$isotherm (right-top), velocity versus $z$ (left-bottom), and the constant temperature perturbation (right-bottom) versus $z$ for the case of a two-dimensionl and open cavity corresponding to the marginal state for $Z_0=0$ and $\lambda=1$}
\label{rcz0open}
\end{figure}

\begin{figure}
    \includegraphics[scale=0.15]{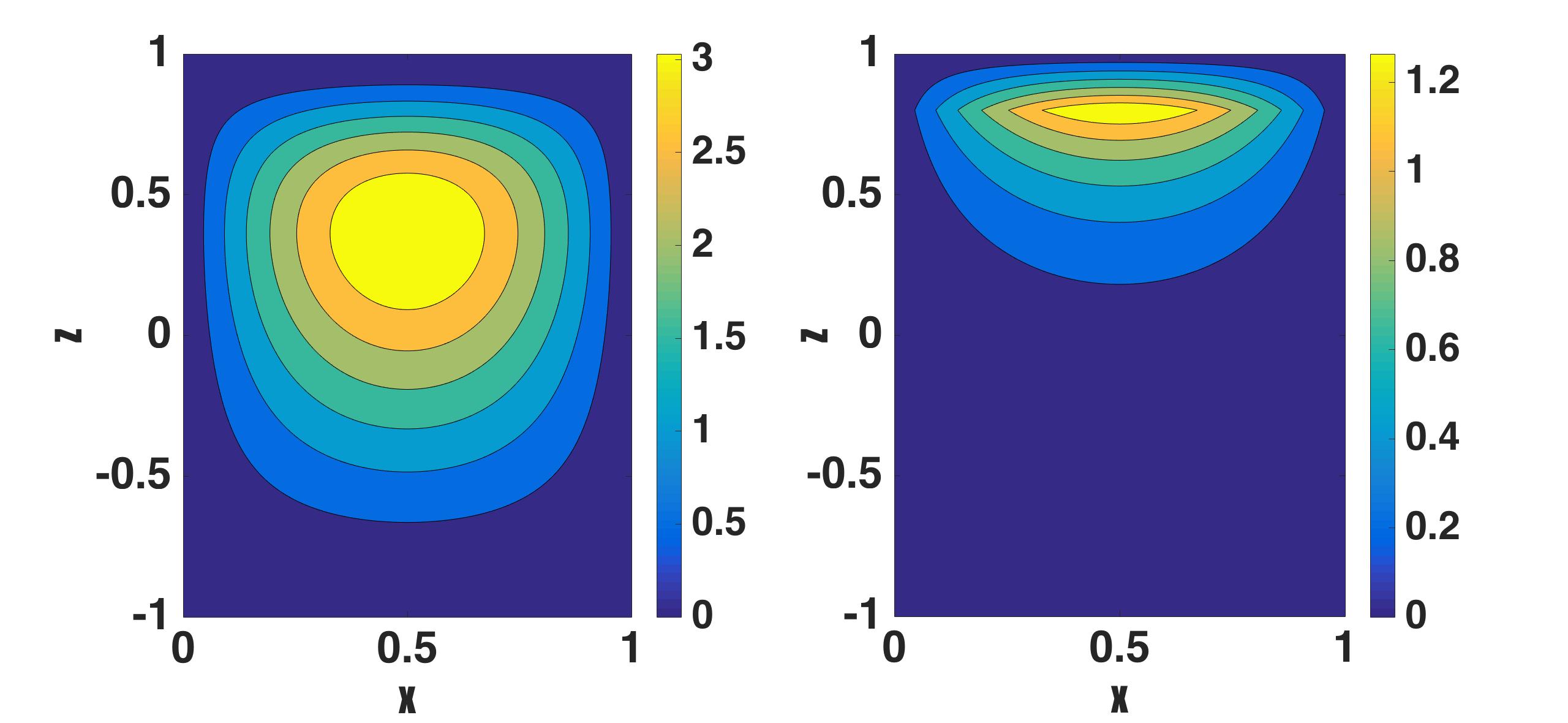}  
    \caption{Plot of the streamlines $\Phi$ (left) and  $\theta$-iso-contours (right) for  $Z_0=0.8$.}
    \label{eyecon}
\end{figure}

 \section{Conclusion}
Buoyancy-driven flows induced by a density stratification whose unstable part is only a fraction of the total fluid height has attracted the attention of several studies due to its relevance to such phenomena as penetrative and astrophysical convection.  A systematic mathematical investigation would lead to systems of equations that are beyond reach for analytical treatment. In this case, it is sometimes beneficial to consider simple idealized models whose solutions offer instructive insight into the stability characteristics of the phenomenon even tough only in a qualitative sense. In this paper, we have investigated Rayleigh-B\'{e}nard convection in both closed and open vertical channels of high aspect ratio when the density distribution in the motionless state has an unstably part that is infinitesimally thin, centrally located and surrounded  from above and below by stably stratified regions. Inspired by the work of Wykes and Dalziel \cite{10}, we have also suggested a way to experimentally set-up such a configuration and test the predictions put forth in this paper using quantitative schlieren technique \cite{11}. We have considered a geometry consisting of a two-dimensional cavity that is either entirely closed or open at the top and bottom so that the convection process is confined horizontally and that the extent of the penetration of the flow from the unstably stratified region into the stable regions can be better appraised. We computed the stability threshold conditions and found them much smaller than those corresponding to a continuous stratification of density in a box \cite{12}. The marginal stability state is attained when the locus of the density discontinuity is at the mid-plane of the layer. This is due to the symmetry in the boundary conditions imposed at the horizontal walls or at $\lambda = \pm \infty$ for the open cavity. The most intense flow is concentrated in the region near the locus of the discontinuity with streamlines having a more elliptical shape than the near-square shape that is encountered in the continuous density stratification case \cite {12}. The temperature, and by the same token the density, have iso-contours that have the shape of symmetric lens whose occurrence is attributed to the jump in the heat flux at the locus of the density discontinuity. Asymmetry in boundary conditions at the vertical walls leads to a critical Rayleigh  number occurring at $Z_0=0$, with flow patterns as depicted in Fig. (~\ref{eyecon}). The streamlines indicate a strong flow near $z=Z_0$ with the less intense flow extending to the lower part of the cell. The iso-contours for the temperature perturbation show an asymmetric lens shape with the largest change occurring at $z=Z_0$. The choice of a thin vertical channel allows for the use of a short membrane so that there is minimal mechanical disturbance to the fluid upon its withdrawal. The experimental testing of the present results hinges on being able to perform the suggested experiment which requires the use of a membrane that is both thin and of very low thermal conductance.

\newpage

\section*{Appendix}

\noindent
The following is the homogeneous system $\hat{D}U=0$
\begin{equation*}\left[\begin{array}{cccccc}
d_{11} & d_{12} & d_{13} &d_{14}&d_{15}& d_{16}\\
d_{21} & d_{22} & d_{23} &d_{24}&d_{25}& d_{26}\\
d_{31} & d_{32} & d_{33} &d_{34}&d_{35}& d_{36}\\
d_{41} & d_{42} & d_{43} &d_{44}&d_{45}& d_{46}\\
d_{51} & d_{52} & d_{53} &d_{54}&d_{55}& d_{56}\\
d_{61}+R q_{61} & d_{62}+Rq_{62} & d_{63}+Rq_{63} &d_{64} & d_{65}& d_{66}\end{array} \right] \left[\begin{array}{c}
B_1\\
B_2\\
B_3\\
C_1\\
C_2\\
C_3\end{array} \right]=\left[\begin{array}{c}
0\\
0\\
0\\
0\\
0\\
0\end{array} \right]\end{equation*}
We can write det$(\hat{D})$  as det$(\hat{D})=$det$(D_1)+$R det$(D_2)$ where 
$$D_1=\left[\begin{array}{cccccc}
d_{11} & d_{12} & d_{13} &d_{14}&d_{15}& d_{16}\\
d_{21} & d_{22} & d_{23} &d_{24}&d_{25}& d_{26}\\
d_{31} & d_{32} & d_{33} &d_{34}&d_{35}& d_{36}\\
d_{41} & d_{42} & d_{43} &d_{44}&d_{45}& d_{46}\\
d_{51} & d_{52} & d_{53} &d_{54}&d_{55}& d_{56}\\
d_{61} & d_{62} & d_{63} &d_{64} & d_{65}& d_{66}\end{array} \right] \text{and}~ D_2=\left[\begin{array}{cccccc}
d_{11} & d_{12} & d_{13} &d_{14}&d_{15}& d_{16}\\
d_{21} & d_{22} & d_{23} &d_{24}&d_{25}& d_{26}\\
d_{31} & d_{32} & d_{33} &d_{34}&d_{35}& d_{36}\\
d_{41} & d_{42} & d_{43} &d_{44}&d_{45}& d_{46}\\
d_{51} & d_{52} & d_{53} &d_{54}&d_{55}& d_{56}\\
q_{61} & q_{62} & q_{63} &0 & 0& 0\end{array} \right].$$
The elements of the matrix $D_1$ and $D_2$ where $l=\pi$ as follows:
\begin{align*}
d_{11}&=-l \lambda \cosh(l (\lambda))+\sinh(l (\lambda))\\
d_{12}&=\lambda \sinh(l (\lambda))\\
d_{13}&=\lambda^2 \sinh(l (\lambda))\\
d_{14}&=-l \lambda \cosh(l (\lambda))+\sinh(l (\lambda))\\
d_{15}&=-\lambda \sinh(l (\lambda))\\
d_{16}&=\lambda^2\sinh(l (\lambda))\\
d_{21}&=-(l)^2 \lambda \sinh(l (\lambda))\\
d_{22}&=\lambda l \cosh(l (\lambda))+\sinh(l\lambda)\\
d_{23}&=\lambda^2l\cosh(l (\lambda))+2\lambda\sinh(l\lambda)\\
d_{24}&=l^2 \lambda \sinh(l (\lambda))\\
d_{25}&=\lambda l \cosh(l (\lambda))+\sinh(l\lambda)\\
d_{26}&=-\lambda^2l\cosh(l (\lambda))-2\lambda\sinh(l\lambda)
\end{align*}
\begin{align*}
d_{31}&=-\lambda l^3 \cosh(l\lambda)-\lambda^2\sinh(l\lambda)\\
d_{32}&=2l\cosh(l\lambda)+l^2\lambda\sinh(l\lambda)\\
d_{33}&=(2+l^2\lambda^2)\sinh(l\lambda)+4l\lambda\cosh(l\lambda)\\
d_{34}&=-\lambda l^3 \cosh(l\lambda)-\lambda^2\sinh(l\lambda)\\
d_{35}&=-2l\cosh(l\lambda)-l^2\lambda\sinh(l\lambda)\\
d_{36}&=(2+l^2\lambda^2)\sinh(l\lambda)+4\lambda l\cosh(l\lambda)\\
d_{41}&=-2l^3\cosh(l\lambda)-l^4\lambda \sinh(l\lambda)\\
d_{42}&=l^3\lambda\cosh(l\lambda)+3l^2\sinh(l\lambda)\\
d_{43}&=(6l+l^3\lambda^2)\cosh(l \lambda)+6\lambda l^2\sinh(l\lambda)\\
d_{44}&=2l^3\cosh(l\lambda)+l^4\lambda \sinh(l\lambda)\\
d_{45}&=l^3\lambda\cosh(l\lambda)+3l^2\sinh(l\lambda)\\
d_{46}&=-(6l+l^3\lambda^2)\cosh(l\lambda)-6\lambda l^2\sinh(l\lambda)\\
d_{51}&=-l^5\lambda\cosh(l\lambda)-3l^4\sinh(l\lambda)\\
d_{52}&=4l^3\cosh(l\lambda)+l^4\lambda \sinh(l\lambda)\\
d_{53}&=8l^3\lambda\cosh(l\lambda)+12l^2\sinh(l\lambda)+l^4\lambda\sinh(l\lambda)\\
d_{54}&=-l^5\lambda\cosh(l\lambda)-3l^4\sinh(l\lambda)\\
d_{55}&=-4l^3\cosh(l\lambda)-l^4\lambda \sinh(l\lambda)\\
d_{56}&=8l^3\lambda\cosh(l\lambda)+12l^2\sinh(l\lambda)+l^4\lambda\sinh(l\lambda)\\
d_{61}&=-4l^4\lambda\cosh(l\lambda)-3l^6\lambda\sinh(l\lambda)\\
d_{62}&=\lambda l^5\cosh(l\lambda)+5l^4\sinh(l\lambda)\\
d_{63}&=(20 l^3+\lambda l^5)\cosh(l\lambda)+5l^4\sinh(l\lambda)\\
d_{64}&=4 l^4\lambda\cosh(l\lambda)+3(n\pi)^6\lambda\sinh(l\lambda)\\
d_{65}&=\lambda l^5\cosh(l\lambda)+5 l^4\sinh(l\lambda)\\
d_{66}&=-(20l^3+\lambda l^5)\cosh(l\lambda)-5 l^4\sinh(l\lambda)\\
q_{61}&=-\lambda l^3\cosh(l\lambda)+l^2\sinh(l\lambda)\\
q_{62}&=l^2\lambda\sinh(l\lambda)\\
q_{63}&=l^2\lambda^2\sinh(l\lambda)
\end{align*}

\end{document}